\documentclass{WileyMSP-template}
\usepackage{xcolor,amsmath}
\usepackage{hyphenat}

\begin{document}


\title{Stiff and deformable quasicrystalline architected materials}

\maketitle
\justifying


\author{Matheus I. N. Rosa,*}
\author{Konstantinos Karapiperis,}
\author{Kaoutar Radi,}
\author{Dennis M. Kochmann*}

\dedication{}

\begin{affiliations}
Dr. M. I. N. Rosa, Dr. K. Radi, Prof. D. M. Kochmann\\
Department of Mechanical and Process Engineering\\
ETH Z\"{u}rich\\
Zurich, Switzerland\\
Email Address: minguaggiato@ethz.ch, dmk@ethz.ch\\
Prof. K. Karapiperis \\
School of Architecture, Civil and Environmental Engineering\\
EPFL\\
Lausanne, Switzerland
\end{affiliations}

\keywords{Architected Material, Metamaterial, Quasicrystal, Mechanical Properties, 3D Printing}

\begin{abstract}
Architected materials achieve unique mechanical properties through precisely engineered microstructures that minimize material usage. However, a key challenge of low-density materials is balancing high stiffness with stable deformability up to large strains. Current microstructures, which employ slender elements such as thin beams and plates arranged in periodic patterns to optimize stiffness, are largely prone to instabilities, including buckling and brittle collapse at low strains. This challenge is here addressed by introducing a new class of aperiodic architected materials inspired by quasicrystalline lattices. Beam networks derived from canonical quasicrystalline patterns, such as the Penrose tiling in 2D and icosahedral quasicrystals in 3D, are shown to create stiff, stretching-dominated topologies with non-uniform force chain distributions, effectively mitigating the global instabilities observed in periodic designs. Numerical and experimental results confirm the effectiveness of these designs in combining stiffness and stable deformability at large strains, representing a significant advancement in the development of low-density metamaterials for applications requiring high impact resistance and energy absorption. Our results demonstrate the potential of deterministic quasi-periodic topologies to bridge the gap between periodic and random structures, while branching towards uncharted territory in the property space of architected materials.
\end{abstract}


\section{Introduction}

The development of lightweight, load-bearing material systems with robust mechanical properties is essential for a wide range of engineering applications. In recent decades, there has been a significant surge in research within the field of architected materials, also known as mechanical metamaterials~\cite{surjadi2019mechanical}. These materials hold the promise of expanding the material property space by utilizing precisely engineered microstructures that require only a small volume fraction of the parent material. Beyond exceptional stiffness and strength~\cite{schaedler2011ultralight,zheng2014ultralight}, their unique properties have also been exploited for diverse applications ranging from the attenuation of sound and vibrations~\cite{liu2000locally,bilal2018architected}, impact resistance~\cite{portela2021supersonic}, biomedical implants~\cite{yavari2015relationship}, and speech recognition~\cite{dubvcek2024sensor}, to name a few. 

For low-density materials a significant challenge lies in achieving high stiffness while maintaining stable deformability under large strains. Low-density topologies typically incorporate slender structural elements such as thin beams~\cite{schaedler2011ultralight,zheng2014ultralight}, plates~\cite{berger2017mechanical,tancogne20183d} or shells~\cite{han2015new,bonatti2019smooth}, which are prone to buckling instabilities when arranged in stiffness-optimal configurations that support loads trough tensile/compressive force chains. For instance, stretching-dominated truss lattices carry loads primarily through uniaxial tension and compression of its members, which makes them orders of magnitude stiffer than bending-dominated counterparts that undergo primarily bending deformation~\cite{ashby2006properties}. However, these stiff, periodic arrangements also experience highly unstable deformation beyond a peak stress, which marks the onset of buckling or crushing and which usually occurs at low compressive strains, leading to catastrophic failure in load-controlled applications. Moreover, periodic designs often lead to anisotropic material behavior, rendering the stiffness and strength highly directional dependent.

Advances to mitigate instabilities have been made, e.g., by considering hollow and hierarchical lattices~\cite{lakes1993materials,meza2015resilient,zheng2016multiscale}, density-graded designs~\cite{niknam2020graded}, spatially variant beam lattices~\cite{maurizi2022inverse}, or complex 3D geometries obtained via topology optimization~\cite{wang2023non}. Although such approaches have led to improvements in buckling strength, they have not fully eliminated the instabilities that plague stiff low-density metamaterials. This behavior renders these materials especially unsuitable for applications requiring large deformation, such as impact mitigation and energy absorption. Here, flexible, bending\hyp{}dominated designs with more stable deformation (i.e., without pronounced softening beyond buckling) are typically preferred~\cite{ashby2006properties}. Stable large-strain deformation has been demonstrated in tensegrity lattices, which leverage discontinuous compressive force chains~\cite{bauer2021tensegrity}. Yet, this comes at a large stiffness sacrifice and requires the fabrication of complex assemblies with pre\hyp{}deformed members. Another promising alternative are interpenetrating lattices, whose architectures are more deformable than the individual constituents---yet again at the expense of stiffness \cite{white2021interpenetrating}. Therefore, the reconciliation of stiffness and deformability within a single material platform remains an important open challenge, which, if solved, would enable stiff, lightweight materials for large-deformation scenarios.

To address this challenge, we have designed, simulated, fabricated, and tested a new class of aperiodic architected materials inspired by quasicrystalline lattices (\textbf{Figure 1}). While they lack translational periodicity, quasicrystals (or quasi-periodic crystals) retain long-range order and exhibit symmetries which are forbidden in periodic crystals. This includes 5, 7, 8, and 10-fold rotational symmetries in the plane, and icosahedral symmetries in 3D~\cite{walter2009crystallography}. These structures form highly symmetric and deterministic patterns, representing an intermediate state between periodicity and disorder~\cite{morison2022order}. Since their discovery in 1984 by Dan Shechtman~\cite{shechtman1984metallic}, quasicrystals have been a vibrant area of research in materials science, including their synthesis~\cite{janot1994quasicrystals}, natural occurrence~\cite{bindi2009natural}, and electronic properties~\cite{poon1992electronic}. Quasicrystals have also inspired advances in metamaterials research, including the manipulation of light in photonic quasicrystals~\cite{man2005experimental,vardeny2013optics}, the observation of isotropic elasticity in quasi-periodic composites~\cite{beli2021mechanics,wang2020quasiperiodic,imediegwu2023mechanical}, and the manipulation of wave propagation~\cite{chen2020isotropic,beli2022wave} and topological localized modes~\cite{ni2019observation,rosa2021exploring} in acoustic and elastic metamaterials. The integration of aperiodicity into architected materials has also shown promise in enhancing failure resistance, as recently demonstrated, e.g., for crack propagation~\cite{wang2024superior} and compressive crushing~\cite{jung2024aperiodicity} of designs inspired by the newly discovered 2D aperiodic monotile~\cite{smith2023aperiodic}. In spite of these notable contributions, the fundamental deformation mechanisms of architected materials with quasicrystalline patterns have remained poorly understood and out of reach for the creation of stiff, low-density architectures such as the designs presented herein. In particular, our dual lattice designs leverage non-uniform distributions of force chains to effectively mitigate global buckling instabilities and failure patterns commonly observed in state-of-the-art periodic counterparts, while providing considerable and, importantly, isotropic stiffness. Therefore, our results represent not only a new class of architected material but also an important step towards uniting stiffness and deformability under large strains in low-density materials.

\begin{figure}[h!]
  \includegraphics[width=\textwidth]{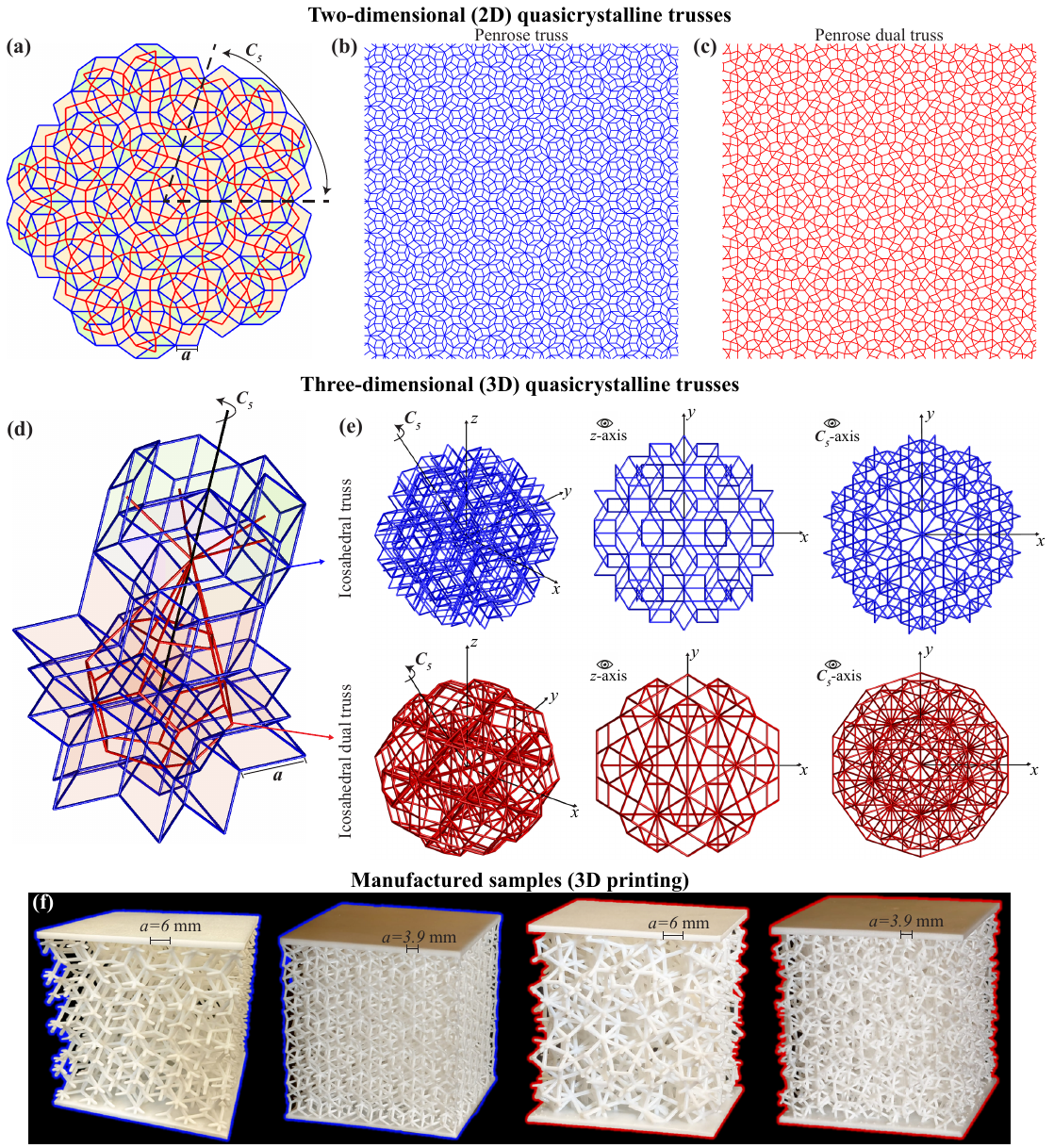}
  \caption{\textbf{Design of quasicrystalline architected materials.} (a) A section of the Penrose tiling, illustrating the formation of (b)~the Penrose truss and (c)~its dual, which are created from the tile edges (blue) and by connecting the centroids of adjacent tiles (red), respectively. (d) The edges and centroids of the 3D icosahedral quasicrystal (IQC) are used to generate the icosahedral truss and its dual (e). (f) Photographs of 3D-printed samples of two different resolutions for the IQC truss (left) and its dual truss (right).}
  \label{fig1}
\end{figure}

\section{Design of quasicrystalline architected materials} 

Our study introduces a novel family of 2D and 3D truss-based architected materials inspired by quasicrystalline structures, as illustrated in Figure~\ref{fig1}. We explore two types of beam networks: one defined by the edges of the quasicrystal tiles (blue) and another one by connecting the centroids of adjacent tiles (red). The latter is referred to as the \textit{dual} truss. The original quasicrystal-based trusses tend to be flexible, primarily characterized by member bending under mechanical loads, while their dual counterparts demonstrate increased stiffness, dominated by axial deformation.

In two dimensions (2D), we focus on the canonical Penrose tiling~\cite{penrose1979pentaplexity} (Figure~\ref{fig1}a), an aperiodic pattern with 5-fold rotational symmetry ($\mathcal{C}_5$) composed of two fundamental tiles of constant edge length $a$. We construct the \textit{Penrose truss} from the network of struts corresponding to the tile edges (blue), while the \textit{Penrose dual truss} is created by connecting the centroids of adjacent tiles (red). Using substitution rules~\cite{grunbaum1987tilings}, we generate tiles of arbitrary sizes, with examples for a square domain of $40a \times 40a$ shown in Figures~\ref{fig1}b,c for the Penrose truss and its dual, respectively. The dual lattice is more complex (containing pentagons, triangles, and other polygons of varying edge lengths), but it maintains the quasi-periodicity and $\mathcal{C}_5$-symmetry of the original tiling.

In three dimensions (3D), we extend this concept to a family of dual trusses derived from icosahedral quasicrystals (IQCs), as depicted in Figure~\ref{fig1}d. These 3D quasicrystals are generated using the cut-and-project method, where the structure is obtained by projecting a six-dimensional hypercube into 3D space~\cite{los1993scaling,chen2020isotropic}. Other methods, such as the multi-grid approach~\cite{levine1986quasicrystals,socolar1986quasicrystals} and substitution rules~\cite{madison2015substitution}, can produce similar quasicrystalline patterns. The shown 3D quasicrystals consist of four types of polyhedra, with edges of equal length $a$ defining the truss structure. The dual truss is constructed by connecting the centroids of adjacent polyhedra, forming a network that retains the quasi-periodicity and symmetries of the original quasicrystal. Detailed design procedures are provided in the Supporting Information (SI) - Note 1. Figure~\ref{fig1}d shows a central patch of the quasicrystal, with the dual truss emanating from a central dodecahedron and comprising beams of varying lengths. Despite the absence of translational periodicity, these 3D quasicrystals are highly ordered and regular, characterized by their unique icosahedral symmetry, which includes 12 axes of $\mathcal{C}_5$-symmetry, 15 axes of $\mathcal{C}_3$-symmetry, and 30 axes of $\mathcal{C}_2$-symmetry~\cite{man2005experimental}. Examples of finite-size IQC trusses and their duals are shown in Figure~\ref{fig1}e, highlighting views along the $z$-axis (exhibiting $\mathcal{C}_2$-symmetry) and along an axis with $\mathcal{C}_5$-symmetry. Figure~\ref{fig1}f presents images of additively manufactured quasicrystal truss samples of two different resolutions, with edge lengths $a=6$~mm and $a=3.9$~mm, which are used in the experiments of this study. Since these trusses lack periodicity, samples were created by truncating larger quasicrystalline structures into cubic domains measuring $5.7\,\text{cm}\times 5.7\,\text{cm}\times 5.7\,\text{cm}$.

\section{Elastic properties} 

To investigate the influence of the quasicrystalline topologies on the elastic properties of architected materials and to compare their fundamental behavior to classic periodic designs, we conducted numerical simulations for both 2D and
3D trusses. (Complementary experiments will be presented in Section~\ref{sec:Experiments}.) Struts are modeled as slender, linear elastic beams that deform according to the classical Euler-Bernoulli theory within a co-rotational frame~\cite{phlipot2019quasicontinuum} to account for geometric nonlinearity such as during buckling and reconfiguration. We here deliberately employ a linear elastic constitutive law to uncover the fundamental (linear and nonlinear) deformation mechanisms emerging from the quasicrystalline truss architectures without confounding those with the effects of material-dependent inelasticity and failure (which will naturally emerge in experiments).

Due to the lack of translational symmetry in quasicrystalline topologies, a single unit cell simulation with periodic boundary conditions is not possible. Instead, representative finite domains of size $40a \times 40a$ for 2D and $9.5a \times 9.5a \times 9.5a$ for 3D are considered, where $a$ represents the fundamental length associated with the edge size of the quasicrystal tiles or the unit cell size in periodic lattices. Motivated by the experimental efforts of this work, we chose a non-multiple of $a$ in 3D designs to avoid beams with half or quarter cross sections at the boundaries of periodic designs. Displacement-controlled uniaxial compression boundary conditions are applied along the vertical direction, and key quantities such as effective compressive stiffness for small (linear) strains and the complete stress-strain response for nonlinear deformation are extracted (see Methods for details).

\begin{figure}[h!]
  \includegraphics[width=\textwidth]{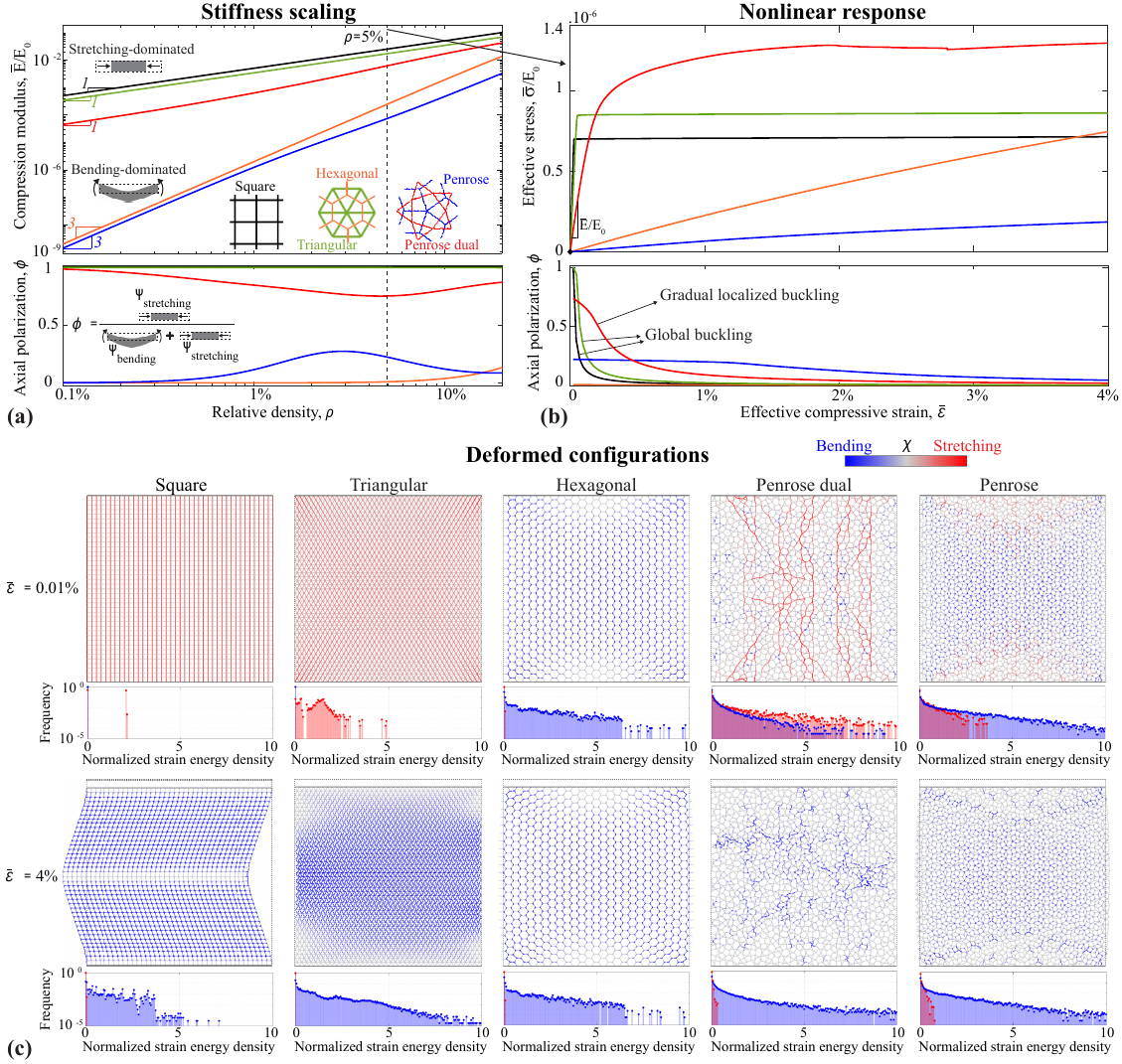}
  \caption{\textbf{Elastic properties of 2D quasicrystalline lattices.} (a) Stiffness scaling with relative density (or fill fraction), highlighting the bending-dominated character of the Penrose lattice (similar to the hexagonal lattice), and stretching-dominated character of its dual (similar to square and triangular lattices). (b) Nonlinear elastic response of lattices at $\rho=5\%$, contrasting the global buckling experienced by square and triangular lattices with the gradual, localized buckling experienced by the Penrose dual lattice. (c) Deformed configurations at two different strain levels and the corresponding distributions of strain energy density, illustrating the (uniform vs.\ localized) distribution of load within the architected materials.}
  \label{fig2}
\end{figure}

The simulated properties of the 2D quasicrystalline trusses are summarized in \textbf{Figure~\ref{fig2}} in comparison with classical periodic topologies, including square, hexagonal, and triangular lattices. Figure~\ref{fig2}a shows the scaling of the effective compressive stiffness with relative density, highlighting the fundamental deformation mechanisms associated with bending and stretching of the struts. As expected for 2D trusses~\cite{fleck2010micro}, the compressive stiffness scales linearly with density for stiff, stretching-dominated designs (such as the square and triangular lattices) and cubically for flexible, bending-dominated ones (such as the hexagonal lattice). The quasicrystal trusses define a pair of stiff and compliant topologies: the stiffness of the Penrose lattice, a bending-dominated design, scales cubically with density, while the Penrose dual lattice, a stretching-dominated truss, shows linear stiffness scaling with density. The hexagonal and triangular lattices also form a pair of compliant and stiff trusses that are dual in the same sense as the quasicrystal tile definition (as illustrated in the insets). The triangular lattice is known for being a stiffness-optimal network~\cite{gurtner2014stiffest}, i.e., it exhibits the maximum isotropic stiffness achievable by 2D beam-based topologies. This makes it an important benchmark, since the quasicrystals studied here are also isotropic (see the SI Figure S4). In comparison, the Penrose dual truss is stiff but does not reach the isotropic limit defined by the triangular truss, which is itself less stiff than the anisotropic square lattice when loaded along the vertical direction considered herein. Note that with increasing density, the calculated stiffness ultimately cease to exhibit constant scaling with relative density, a behavior previously reported for 3D periodic trusses~\cite{meza2017reexamining} and rationalized by the growing influence of beam junctions \cite{PortelaEtAl2018}.

To further characterize the deformation mechanisms, the strain energy is separated into bending and axial (stretch/compression) components during loading. An axial polarization coefficient, $\phi$, is defined as the ratio of axial strain energy to total strain energy (see Methods). This coefficient approaches $1$ for deformation modes dominated by axial stretching and compression, and $0$ for bending dominance. The variation of $\phi$ with relative density, included in Figure~\ref{fig2}a for the linear regime, reveals the stretching and bending-dominated nature of the designs. The Penrose lattice exhibits a purely bending-dominated response at low relative densities but shows a progressive increase in $\phi$, indicating a small degree of axial deformation (the maximum $\phi=0.275$ is at around $\rho=2.5\%$). Similarly, the Penrose dual lattice, which is mostly stretching-dominated starting from $\phi=1$ at low $\rho$, reaches a minimum of $\phi=0.75$ at around $\rho=5\%$, indicating a small degree of bending deformation. Interestingly, these variations are much more pronounced for the quasicrystal lattices than in the classic truss topologies. (Note that the exact values of $\phi$ may depend on sample size and boundary conditions, so that these results do not necessarily correspond to the homogenized properties of infinite domains, whose identification is a challenging task for the non-periodic quasicrystals.)

The stress-strain responses of the lattices under nonlinear deformation is presented in Figure~\ref{fig2}b for $\rho=5\%$, while Figure~\ref{fig2}c illustrates the deformed shapes of samples for $\bar\epsilon=0.01\%$ (linear regime) and $\bar\epsilon=4\%$ (final strain in the simulation). The hexagonal and Penrose lattices exhibit smooth, compliant responses characteristic of their bending-dominated nature, with nonlinearities arising primarily from the reconfiguration of struts. Conversely, the square, triangular, and Penrose dual lattices are stiffer and prone to buckling instabilities due to their stretching-dominated nature. A key finding of this work is the markedly different manifestation of buckling instabilities in periodic versus quasi-periodic lattices. The triangular and square lattices experience global buckling instabilities at a low strain of about $\bar\epsilon=0.05\%$, followed by an almost constant stress plateau. In contrast, the Penrose dual lattice primarily undergoes local instabilities, as global buckling modes cannot form due to the lack of periodicity. This results in a smooth, stable stress-strain response characterized by a cascade of sequential local instabilities, reaching a plateau at a strain of about $\bar\epsilon=2\%$, which is $40$ times higher than the buckling strain of the square and triangular lattices and occurs at a higher stress level. This behavior is further evidenced by the evolution of the axial polarization index $\phi$ as a function of strain accompanying the stress-strain plots (included in Figure~\ref{fig2}b). Initially high $\phi$ values, indicative of stretching-dominated deformation, inevitably degrade with increasing strains due to buckling instabilities and a transition toward bending-dominated deformation in the post-buckled regime. In periodic lattices, $\phi$ decreases rapidly due to global instabilities, whereas the Penrose dual lattice exhibits a smoother transition due to distributed local instabilities.

Global buckling instabilities arise from uniform force chains in periodic stretching-dominated lattices, which do not manifest in quasi-periodic designs, as illustrated in Figure~\ref{fig2}c. The deformed configuration plots are color-coded according to an energy indicator, $\chi$ (see Methods), whose magnitude equals the strain energy density, while the sign is used to visually differentiate members under tension/compression (red) from those undergoing bending (blue). Each plot is accompanied by a histogram showing the distribution of energy density components across all struts, normalized by the mean value. In the linear regime ($\epsilon=0.01\%$), the square and triangular lattices exhibit predominantly axial strain energy, forming uniformly distributed force chains throughout the domain, reflected in concentrated histogram distributions. While such uniform loading imparts high stiffness, it also leads to global buckling instabilities, as shown in the lower plots of Figure~\ref{fig2}c ($\epsilon=4\%$). The square lattice is loaded primarily through vertical columns of force chains aligned with the compression direction, resulting in an almost Dirac-like distribution of axial strain energy and leading to the shown global buckling mode, consistent with other studies~\cite{he2018buckling}. The triangular lattice shows a slightly more dispersed concentration of strain energy in the histogram due to boundary effects, but it still results in a global buckling-induced transformation, also consistent with previous observations~\cite{kang2014complex}. In sharp contrast, the Penrose dual lattice exhibits a more intricate pattern of force chains that are not uniformly distributed, reflected in a long-tailed histogram. Its nonlinear response captures a sequence of localized instabilities that prevent instant global collapse. Indeed, the deformed configuration for $\epsilon=4\%$ shows multiple buckled elements (evidenced by high bending energy), but a large portion of the domain still exhibits moderate and low strain energies without buckling. This suggests a promising failure resistance in quasicrystalline designs, which retain high stiffness while avoiding global buckling. Finally, both the hexagonal and Penrose lattices exhibit compliant responses characterized primarily by bending strain energy. A small amount of stretching strain energy is present in the Penrose lattice in the linear regime, correlating with the non-zero axial coefficient $\phi=0.22$ and emanating from a few force chains near the boundaries. As strain increases, deformation becomes dominated by bending energy, as evidenced by the continuous decrease in $\phi$. 

\begin{figure}[b!]
  \includegraphics[width=\textwidth]{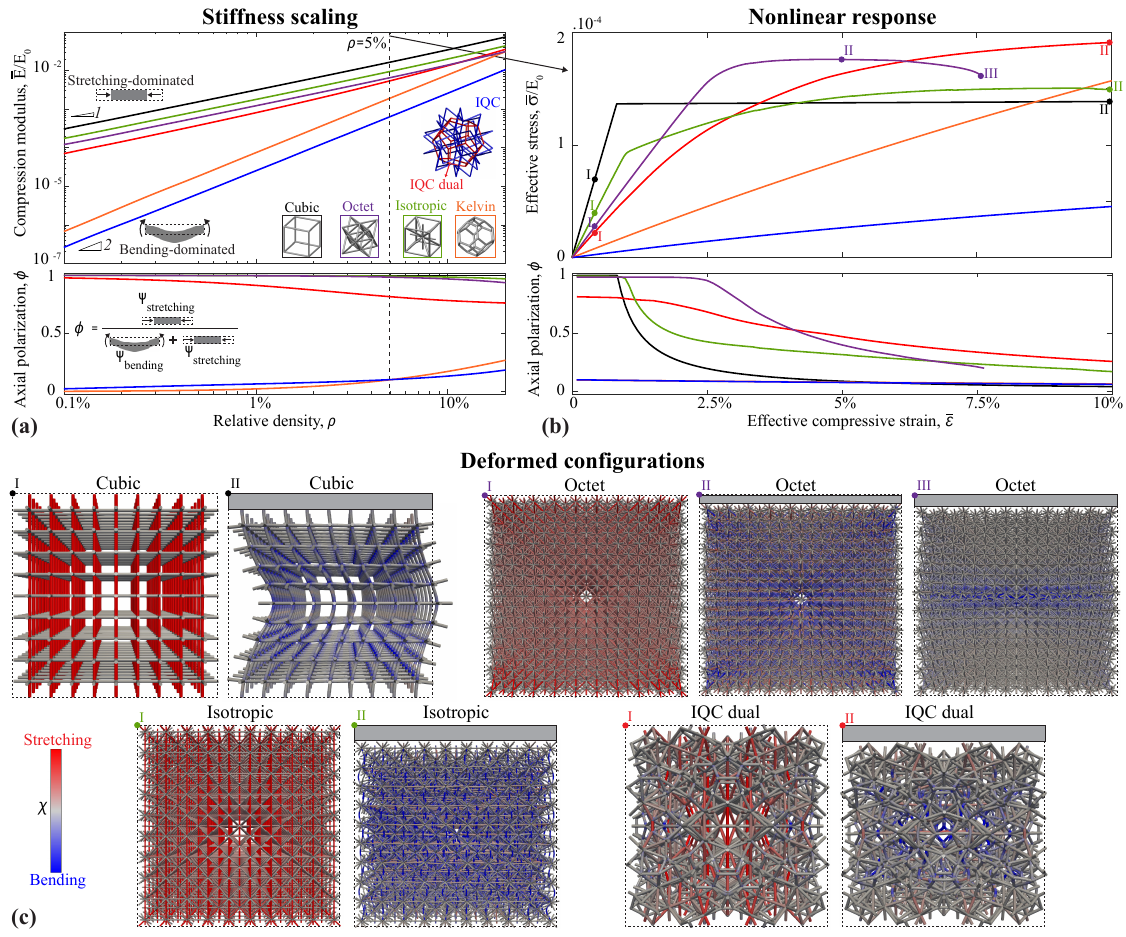}
  \caption{\textbf{Elastic properties of 3D quasicrystalline lattices.} (a) Stiffness scaling with density, highlighting the bending-dominated character of the icosahedral lattice (along with the Kelvin lattice), and stretching-dominated character of its dual (along with the cubic, octet, and isotropic lattices). (b) Nonlinear elastic response of lattices at $\rho=5\%$, contrasting the global buckling experienced by the periodic stretching-dominated lattices and the gradual localized buckling experienced by the IQC dual lattice. Same colors represent the same topologies. (c) Deformed configurations and axial vs.\ bending energy contributions at strain levels marked in (b).}
  \label{fig3}
\end{figure}

These findings qualitatively extend to 3D quasicrystalline designs. \textbf{Figure~\ref{fig3}} compares their properties to periodic topologies such as the simple cubic lattice, chosen for its simplicity and pronounced global buckling characteristics, as well as the octet and Kelvin lattices, which are traditional examples of stretching- and bending-dominated designs, respectively, favored for their enhanced stiffness/strength and energy absorption characteristics~\cite{ashby2006properties}. Additionally, we include the isotropic lattice as an important benchmark, which achieves optimal isotropic stiffness in the context of 3D truss designs~\cite{gurtner2014stiffest}. The variation of the compressive stiffness with relative density (Figure~\ref{fig3}) separates stretching\hyp{}dominated lattices (cubic, octet and isotropic) characterized by a linear scaling with density, from bending\hyp{}dominated lattices (Kelvin), which exhibit quadratic scaling in 3D~\cite{ashby2006properties}. Similarly to the 2D case, the 3D quasicrystalline designs define a pair of stiff and compliant lattices. The IQC truss (blue) is bending\hyp{}dominated and more compliant than the Kelvin lattice, while its dual (red) is stretching\hyp{}dominated but not quite as stiff as the optimal isotropic truss, which in turn is less stiff than the simple cubic lattice (when compressed vertically). However, the stiffness of the dual lattice is comparable to that of the octet lattice and in fact surpasses it at about $\rho=12\%$. The polarization coefficient $\phi$ also confirms their primary deformation mechanisms, with values close to $0$ and $1$ for bending- and stretching-dominated designs, respectively, and variations appearing towards higher relative densities as in the 2D case. 

The nonlinear response of the 3D lattices (Figure~\ref{fig3}b) mirrors the trends observed in~2D. Bending-dominated trusses exhibit a compliant and stable behavior with reconfiguration-induced nonlinearity, whereas stretching-dominated lattices are prone to buckling instabilities. The deformed shapes (Figure~\ref{fig3}c) further illustrate this contrast: periodic designs, with their uniform distribution of force chains, tend to experience global instabilities, while the dual IQC lattice, characterized by non-uniformly distributed force chains, undergoes mostly localized instabilities. Additional insight into the nature of these force chains, as revealed by the distribution of axial vs.\ bending strain energy, are provided in the SI (Note 2, Figure S3). In the cubic lattice, a distinct global buckling shape is observed, marked by a sudden, flat stress plateau. The octet and isotropic trusses, on the other hand, display more complex patterns of buckled beams, resulting in slightly smoother transitions into their stress plateaus. Notably, the octet lattice experiences a second post-buckling instability as deformation becomes concentrated in a collapsed central layer, at which point the simulation is halted to prevent nonphysical results due to the lack of contact modeling. Such localized layer collapse is a well-known, yet undesirable, post-buckling characteristic of stretching-dominated periodic architectures~\cite{ashby2006properties,bauer2021tensegrity}, also observed in the experimental portion of this work. In contrast, the IQC dual lattice exhibits stable deformation with localized buckled beams, which occur in far fewer numbers compared to the periodic lattices, effectively avoiding global instabilities and layer collapses. The evolution of the axial polarization coefficient $\phi$ with strain further highlights a smoother transition in the IQC dual lattice, in stark contrast to the rapid decay observed in the periodic counterparts.

In summary, the dual truss topologies are stiff, stretching-dominated structures that form non-uniformly distributed force chains, resulting in relatively stable deformation without global failure. An important additional feature, the considered quasicrystalline designs are elastically isotropic due to the large number of symmetry axes (in line with previous studies of the homogenized linear elastic properties of  quasi-periodic materials~\cite{wang2020quasiperiodic,beli2021mechanics,imediegwu2023mechanical}). Although these designs do not reach the isotropic linear stiffness limit of optimally loaded periodic lattices (triangular in 2D and isotropic truss in 3D), we observe a remarkably high degree of isotropy across a significant strain range in the nonlinear regime (see SI Figure S4). This feature, together with the avoidance of global buckling, greatly enhances their deformability, suggesting the potential for failure-resistant, isotropic, low-density materials capable of absorbing larger amounts of elastic energy while maintaining high stiffness. This potential is further illustrated by the experimental results reported in the following.

\begin{figure}[h!]
  \includegraphics[width=\textwidth]{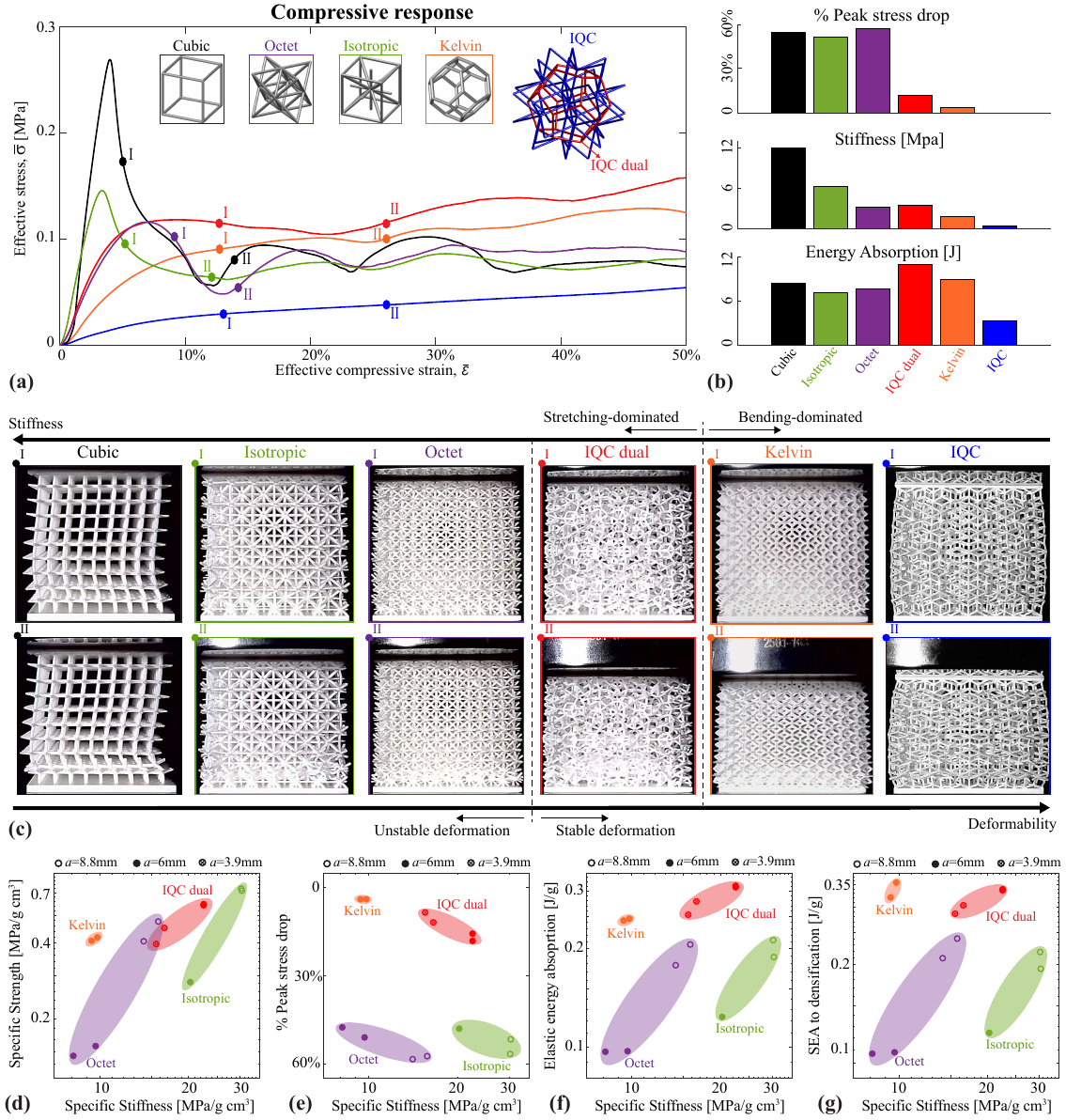}
  \caption{\textbf{Experimental results of uniaxial compression tests on samples with $10\%$ relative density.} (a) Stress-strain responses of selected samples, whose topologies are indicated by the insets. (b) Bar plots showing $\%$ drop in peak stress, stiffness, and energy absorption for each sample. Same colors correspond to the same topologies. (c) Images of deformed samples for selected points marked in (a), highlighting the unique positioning of the icosahedral dual lattice as a stretching-dominated topology with stable response. (d-g) Property space plots correlating stiffness with strength, $\%$ peak stress drop, and specific energy absorption (SEA) up to $50\%$ strain and to densification.}
  \label{fig4}
\end{figure}

\section{Experiments}
\label{sec:Experiments}

\textbf{Figure~\ref{fig4}} reports the results of displacement-controlled uniaxial compression tests conducted on additively manufactured 3D samples. Utilizing a ductile polymer (UMA 90 resin) as the base material allows for the observation of buckling instabilities in large-deformation experiments without triggering brittle collapse. The samples of $10\%$ relative density define cubic domains of side length $a=57$~mm, with the addition of two plates of thickness $2$mm at the top and bottom to facilitate equal load distribution. Representative examples of stress-strain responses up to $50\%$ strain are shown in Figure~\ref{fig4}a, with selected metrics extracted from these curves displayed in Figure~\ref{fig4}b. Images of the samples' deformed shapes at the indicated strain levels are displayed in Figure~\ref{fig4}c. These results were selected out of a larger set of experiments containing two samples of each kind and different $a$-values (Figure~\ref{fig1}f) for a focused comparison against selected stretching-dominated designs. Note that it is, unfortunately, practically impossible to ensure the same number of beams or unit cells and the same $a$-value in a comparison of different periodic and aperiodic architectures. The full range of different $a$-values is detailed in the SI Note~3. Here, we highlight the most important common features, while considering the least favorable case for aperiodic designs as a worst-case scenario to safeguard our observations against a possible length scale bias. The property space plots in Figure~\ref{fig4}d-g include all samples with different $a-$values tested, while the complete stress-strain responses until densification for those are provided in the SI Note~4. 

The experimental measurements (Figure~\ref{fig4}) show good overall agreement with our simulated predictions regarding the post-buckling stability. The bending-dominated lattices, such as the Kelvin and IQC lattices, exhibit compliant responses with lower stiffness and stable deformation even at large strains. In contrast, the stretching-dominated periodic lattices (cubic, octet, and isotropic) exhibit higher linear stiffness but are prone to instability by buckling at low strains. This instability manifests itself in the stress-strain response as a pronounced peak stress followed by an unstable region of negative stiffness, which, although observed in this displacement-controlled setting, would lead to catastrophic failure in a load-induced scenario. The peak stress marks the onset of global buckling instability, rapidly progressing into localized layer collapses. The experimental buckling mode of the cubic lattice closely resembles the simulated one (Figure~\ref{fig3}), whereas higher connectivity lattices like the octet and isotropic lattices show deviations primarily due to imperfections in the sample geometry combined with inelastic deformation (such as yielding and fracture). Concentrating particularly at the nodes during buckling, this promotes layer collapse, as seen in Figure~\ref{fig4}c.

In stark contrast, the stretching-dominated dual IQC lattice exhibits a significantly more stable response while retaining a stiffness comparable to that of the octet lattice. Figure~\ref{fig4}b demonstrates this stability, showing a roughly 55$\%$ drop in peak stress for the cubic, octet, and isotropic samples, compared to only an 11$\%$ drop for the dual IQC lattice. Remarkably, the dual IQC's stability is closer to that of the bending-dominated Kelvin lattice, which exhibits only a $4\%$ drop.  As predicted by numerical simulations, the non-uniform distribution of force chains in the dual IQC lattice leads to distributed failure, effectively preventing global instabilities. Indeed, at a strain of $\bar\epsilon=12\%$, the dual IQC exhibits a deformation pattern with no signs of global buckling or layer collapse (point I in Figure~\ref{fig4}), akin to the behavior observed in the Kelvin and IQC lattices. In contrast, all the periodic stretching-dominated lattices have already undergone global buckling and layer collapse at the same strain (points II on their respective curves). The dual IQC and Kelvin lattices only experience crushing and contact at a larger strain of about $25\%$ (points II), with the Kelvin lattice failing in the bottom layer and the dual IQC lattice showing failure in a more distributed region near the bottom. Additionally, the IQC lattice displays the most stable response, with no detected instability or negative stiffness region, similar to the behavior reported in tensegrity lattices~\cite{bauer2021tensegrity}. In both cases, this stability gain comes at the cost of lower stiffness, even when compared to the Kelvin lattice. The combination of stiffness and stability found in the dual IQC lattice results in a stress plateau that surpasses all other samples, leading to an energy absorption of 11~J up to $50\%$ strain -— 54$\%$ higher than the optimally-stiff isotropic truss and $22\%$ higher than the Kelvin lattice, which has been regarded as an excellent energy absorber~\cite{ashby2006properties}.

The performance of the IQC dual design is further contrasted with periodic lattices in Figures~\ref{fig4}d-g, where specific stiffness is correlated with specific strength, percentage drop in peak stress, and specific energy absorption (SEA). The comparison focuses on the stretching-dominated octet and isotropic lattices at two different length scales, with the Kelvin lattice also included as a state-of-the-art bending-dominated design for energy absorption. (We deliberately exclude the cubic truss from the comparison since its response is extremely direction-dependent, which makes for an unfair comparison.) The IQC dual lattice positions itself competitively in terms of stiffness and strength (Figure~\ref{fig4}d), falling between the isotropic and octet lattices. The isotropic lattice at $a=8.8$~mm has the highest observed stiffness and strength, yet both the isotropic and octet lattices show a sharp decline in these values at the smaller size ($a=6~$mm). This decline is attributed to the buckling sensitivity to imperfections, which is exacerbated in samples with a greater number of beams with thinner cross-sections. In contrast, the IQC dual lattice exhibits a significantly smaller variation with size, with even the worst-case scenario ($a=3.9$~mm) matching the stiffness and strength of the best-performing octet lattices.

The correlation between the percentage drop in peak stress and stiffness (Figure~\ref{fig4}e) highlights the clear advantage of the dual IQC lattice. It occupies the upper right corner of the plot, well-separated from the stretching-dominated counterparts due to its relatively stable deformation and from the Kelvin lattice due to its increased stiffness. This combination also provides significant benefits in terms of combined stiffness and energy absorption (Figures~\ref{fig4}f-g), where the IQC dual lattice also consistently occupies the upper right corner of the plots. Notably, the IQC dual lattice surpasses all other designs in SEA at $50\%$ strain, outperforming the Kelvin lattice (the second-best performer) by 17.5$\%$ on average. Although the Kelvin lattice shows a slight advantage in SEA to densification, averaging 6.8$\%$ higher than the dual IQC lattice due to larger densification strains (see SI Figure S6), this comes at the cost of reduced stiffness. Therefore, the IQC dual lattice continues to occupy the upper right corner of the plot, with significantly higher SEA to densification when compared to the octet and isotropic lattices.

\begin{figure}[b!]
  \includegraphics[width=\textwidth]{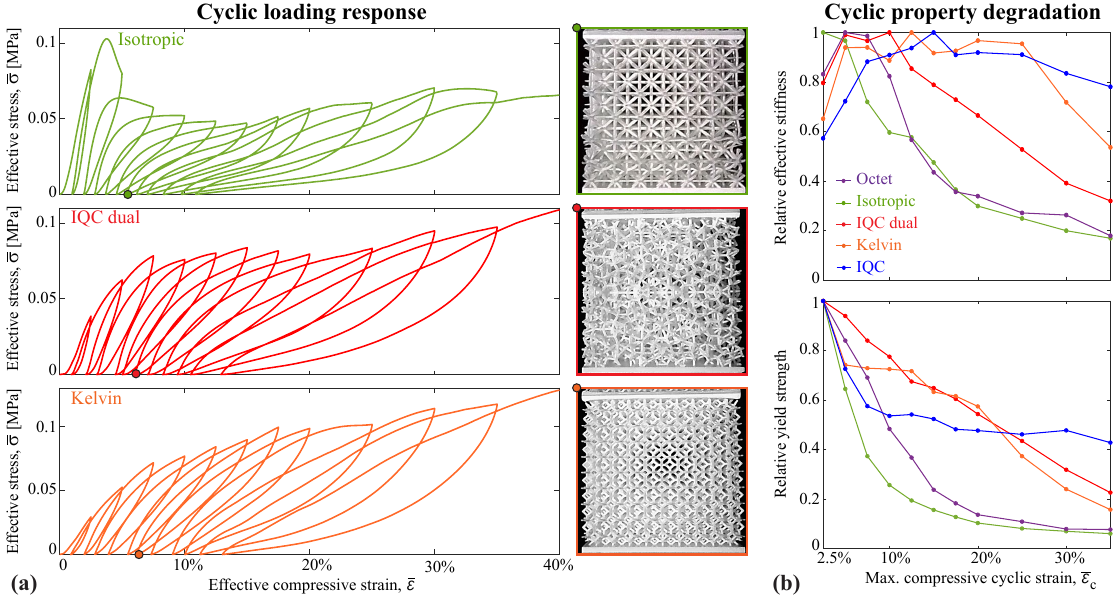}
  \caption{\textbf{Experimental results for cyclic compression tests.} (a) Stress-strain curves contrasting the unstable response of the isotropic lattice with the stable responses of the dual IQC and Kelvin latticess. Photographs show the samples' deformed state after the 6th cycle at an imposed compressive strain of $\bar\epsilon=15\%$. (b) Variation of relative effective stiffness and yield strength with the maximum strain in each cycle.}
  \label{fig5}
\end{figure}

To assess the failure resistance of the proposed designs, we performed displacement-controlled loading-unloading experiments with progressively increasing cyclic strains (\textbf{Figure~\ref{fig5}}a). The results once again underscore the inherent instability of periodic stretching-dominated designs, as evidenced by the isotropic lattice's response, which exhibits significant stress drops throughout the cyclic loading. The accompanying image of the deformed sample, captured after a cyclic strain of $\bar\epsilon=15\%$ during the 6th cycle, reveals substantial damage, including several buckled beams and two collapsed layers. In stark contrast, the dual IQC lattice (whose image was taken at the same strain) demonstrates a considerably more stable response, akin to that of the Kelvin lattice, with no significant stress drops and no signs of global collapse or layer-localized damage in their deformed configurations at the same cyclic strain.

To further illustrate the superior damage resistance of the dual IQC lattice, we present the variation of effective stiffness and yield strength with cyclic strain in Figure~\ref{fig5}b, with values normalized by the maximum observed for each sample. All samples experience degradation of their elastic properties due to damage accumulated during cyclic loading, with some samples showing an initial increase in stiffness before degradation which we attribute to accommodation and potential hardening effects. The octet and isotropic samples exhibit rapid degradation of both stiffness and yield strength, a consequence of their unstable deformation and collapse at low strains. In contrast, the bending-dominated lattices (Kelvin and IQC) show a delayed and more gradual degradation, as expected due to their increased compliance and lack of global buckling instabilities. Notably, the dual IQC lattice displays stable and nearly linear property degradation, with values close to those of the bending-dominated lattices. For example, at a cyclic strain of $\bar\epsilon=15\%$, the IQC lattice retains approximately $80\%$ of its relative stiffness and $65\%$ of its relative yield strength. This is comparable to the Kelvin lattice, which retains $91\%$ and $63\%$, respectively. In stark contrast, the octet and isotropic lattices have already degraded significantly, retaining only about $45\%$ of their relative stiffness and $20\%$ of their relative yield strength. This clear distinction between the periodic stretching-dominated lattices and the dual IQC lattice persists across the entire range of cyclic strains, and it highlights the superior damage resistance of the latter: the dual IQC lattice achieves stable, distributed failure within a stretching-dominated topology, contrasting with the unstable collapse typically seen in periodic stretching-dominated structures, while also presenting higher stiffness than bending-dominated counterparts.

\section{Discussion}

This work has tackled the critical challenge of balancing stiffness and large-strain deformability in architected materials. Traditionally, periodic designs yield either stiff, stretching-dominated lattices prone to buckling instabilities or compliant, bending-dominated lattices that offer a more stable response at the cost of stiffness. To bridge this gap, we have introduced and characterized novel aperiodic architected materials derived from quasicrystalline lattices. Through numerical simulations, we examined their deformation mechanisms, linear stiffness scaling, and the emergence of buckling instabilities in their nonlinear response. Our results indicate that beam-based lattices formed by the edges of the quasicrystal tiles exhibit compliant, bending-dominated behavior, while their dual designs, created by connecting the centroids of adjacent tiles, result in stiff, stretching-dominated topologies. Importantly, these dual lattices (introduced here for the first time) feature non-uniform force chain distributions that simultaneously promote stiffness and stable post-buckling deformation, preventing the unstable collapse often seen in periodic stretching-dominated structures. These characteristics have been validated through experiments demonstrating the unique performance of the dual lattices with relatively high stiffness, superior elastic energy absorption and enhanced failure resistance under cyclic loading. These advantages are expected to be even more pronounced in ultra-low-density materials~\cite{schaedler2011ultralight,zheng2014ultralight}, since the stiffness of bending-dominated designs degrades drastically at lower densities (due to a higher scaling exponent), while periodic stretching-dominated designs remain largely unstable. Moreover, the quasi-periodic designs showed less sensitivity to imperfections due to the absence of translational symmetry, which, again, plays a significant role at smaller scales.

Despite the promising results presented here, several key areas remain to be explored, including the influence of size and boundary conditions, their detailed sensitivity to defects~\cite{shaikeea2022toughness,ziemke2024defect}, and their performance under dynamic impact~\cite{butruille2024decoupling}. Additionally, our findings are based on a small number of non-optimized quasicrystal designs, which already demonstrate advantages over their periodic counterparts. A much broader and largely unexplored design space for quasi-periodic architectures still remains open for investigation. This space includes alternative truss designs, potentially derived from other quasi-periodic tiles~\cite{imediegwu2023mechanical,smith2023aperiodic} or by variations in projection rules from higher-dimensional periodic spaces, further modifiable by the incorporation of hollow tube cross-sections~\cite{meza2015resilient}, or the combination with structural optimization. Beyond beam-based designs, other quasicrystal-derived topology types such as plates~\cite{berger2017mechanical,tancogne20183d} or curved shells~\cite{han2015new,bonatti2019smooth} also warrant exploration.

Our findings indicate that deterministic quasi-periodic topologies hold great promise in bridging the gap between periodic and aperiodic architected materials, thereby enabling the exploration of previously uncharted regions of the material property space. The growing interest in the inverse design of architected materials with tailored properties, achieved through, e.g., topology optimization~\cite{sigmund2009systematic,osanov2016topology} or machine learning~\cite{zheng2023deep,bastek2022inverting}, underscores the significance of this area. Current approaches predominantly focus on periodic designs and their variations, often overlooking the potential of quasi-periodic designs, such as those presented here. The latter offer new opportunities in the material property space, particularly those associated with large nonlinear deformation. Our study demonstrates the potential to balance stiffness and deformability under substantial strains, a critical characteristic for low-density materials in applications requiring large deformations, such as elastic energy absorption and impact mitigation. With improved mechanical performance, such designs also become attractive for applications requiring both mechanical integrity and other key physical phenomena, such as the confinement of electromagnetic~\cite{man2005experimental,vardeny2013optics}, acoustic~\cite{ni2019observation} or elastic~\cite{rosa2021exploring} waves, and the control of thermal properties like heat conductivity~\cite{meyer2022graph}.


\section*{Methods}
\threesubsection{Numerical Simulations}\\
Finite element simulations employ a co-rotational beam model that assumes small local strains but allows for large displacements and rotations~\cite{crisfield1990consistent}. Beams in 2D have a rectangular cross-section of unitary out-of-plane thickness $h$ and an in-plane width $w$ which is dictated by the relative density $\rho$ through the approximate relation $w=\rho {L^2}/{(\sum l_b)}$, where the summation of beam lengths $l_b$ is carried over all the elements within a square domain of size $L=40a$. In 3D, beams have circular cross sections, whose radii are also dictated by the relative density but through more precise relationships established by 3D CAD models (see SI Note 3). All beams have a Young's modulus $E_0$ and the same cross-section within a given structure, with the only exception being the 3D isotropic truss, which requires a ratio of $1.14$ between the radii of the diagonal and non-diagonal struts~\cite{gurtner2014stiffest}. We enforce displacement\hyp{}controlled boundary conditions by specifying the vertical displacement of the nodes on the top boundary of the domains, while setting to zero all remaining degrees of freedom (displacements and rotations) of nodes on both the top and bottom boundaries. We define the effective strain $\bar\epsilon$ as the ratio between the imposed displacement and the vertical height of the domain, while the effective stress $\bar\sigma$ is calculated by summing the vertical reaction forces on the top nodes and dividing by the perpendicular area, i.e. $\bar\sigma=\sum(F)/A$ with $A=hL$ in 2D and $A=L^2$ in 3D. The effective compressive modulus is obtained for small imposed displacements in the linear regime as $\bar E=\bar\sigma/\bar\epsilon$, while the nonlinear response is estimated by incrementing $\bar\epsilon$ in small steps and seeking equilibrium through a Newtown-based line search algorithm. Small and randomized perturbations of magnitude $\delta=0.001a$ are applied to the initial nodal positions of all structures to facilitate the solution stepping through buckling instabilities. The strain energy density $\Psi$ of a beam is decomposed into its axial and bending components, with $\Psi_\text{axial}=E_0\epsilon_{l}^2/2$, where $\epsilon_l$ is the local axial strain of the neutral line, and $\Psi_\text{bending}=\Psi-\Psi_\text{axial}$. The axial polarization coefficient is defined as $\phi=\sum(l_b A_b\Psi_\text{axial})/\sum(l_b A_b \Psi)$, where the summation considers all beams in the domain with length $l_b$ and cross-sectional area $A_b$. The energy indicator is defined for each beam as $\chi=\Psi\cdot\textrm{sign}(\Psi_\text{axial}-\Psi_\text{bending})$, which has the same magnitude as $\Psi$ and a positive sign if $\Psi_\text{axial}>\Psi_\text{bending}$, and a negative sign otherwise. 

\threesubsection{Fabrication}\\
Specimens were fabricated through digital light synthesis (DLS) using a Carbon M2 3D printer with average writting speed of 30mm/hr and layer thickness of 100$\mu$m. The one-part UMA 90 resin was used as the printing material, having a tensile modulus of $484$~MPa, ultimate strength of 12~MPa, and 20$\%$ elongation at break. After printing, samples were immersed in isopropyl alcohol (IPA) on an orbital shaker for 5 minutes to remove excess resin, followed by 4 minutes of post-curing in a UV chamber to ensure complete cross-linking of polymer chains. The relative density of samples with different topologies and sizes was computed with the help of 3D CAD models (see the SI Note 3 for details).  

\threesubsection{Experimental characterization}\\
Uniaxial displacement-controlled compressive experiments were performed at a quasistatic strain rate of 0.00005s$^{-1}$, using an Instron 5943 Single Column testing machine. The effective strains and stresses were determined by the recorded force-displacement curves and the measured sample dimensions. The effective stiffness for each experiment was determined as the maximum slope in the linear elastic regime. Where applicable, the ultimate strength was determined as the maximum recorded stress before the onset of softening. The energy absorption during deformation was computed by integrating the stress-strain curves until 50$\%$ strain or until densification (as indicated in the figures). The densification strain was determined as the global maximum of energy absorption efficiency of the stress-strain responses~\cite{li2006compressive}. The measure of peak instability ($\%$ peak drop) was determined as $(\bar\sigma_\text{max}-\bar\sigma_\text{min})/\bar\sigma_\text{max}$, where $\bar\sigma_\text{max}$ and $\bar\sigma_\text{min}$ correspond to the first local maximum and minimum of the stress-strain curves, respectively. Cyclic loading experiments were performed at the same quasistatic strain rate, with each cycle loaded up to an incrementally increased maximum strain followed by unloading until the force reaches zero. Tests started with an imposed strain of $\bar\epsilon=2.5\%$ and were incremented by $2.5\%$ in each cycle up to $20\%$, after which the increment was increased to $5\%$ until $40\%$. The effective stiffness was determined as the maximum slope of the linear elastic region at the beginning of each cycle, while the yield strength is defined as the $0.2\%$ strain offset relative to the linear effective stiffness~\cite{bauer2021tensegrity}.

\medskip
\textbf{Supporting Information} \par 
Supporting Information is available from the Wiley Online Library.

\medskip
\textbf{Acknowledgements} \par 

M.I.N.~Rosa acknowledges Dr.~Paul J.~Steinhardt, Dr.~Joshua Socolar, and Dr.~Alexey E. Madison for insightful comments on the construction of 3D quasicrystalline lattices. M.I.N.~Rosa also extends gratitude to Elias Pescialli for his support in setting up the cyclic loading experiments, and to Jung-Chew Tse for his assistance with 3D printing.

\medskip

%
\bibliographystyle{MSP}
\bibliography{References}


\end{document}